\pdfoutput=1

\documentclass[11pt,twoside,a4paper,cmspaper,final,collab]{cms-tdr}
\def\svnVersion{60f0289-D}\def\svnDate{2020/03/05}\def\cmsCernNoTag{CERN-EP-2019-255}\def\cmsCernDate{\today}\def\cmsMessage{"Published in the Journal of High Energy Physics as \href{http://dx.doi.org/10.1007/JHEP02(2020)191}{\doi{10.1007/JHEP02(2020)191}.}"}

\begin{document}\cmsNoteHeader{TOP-18-005}

\hyphenation{had-ron-i-za-tion}
\hyphenation{cal-or-i-me-ter}
\hyphenation{de-vices}
\RCS$Revision: 494881 $
\RCS$HeadURL: svn+ssh://otoldaie@svn.cern.ch/reps/tdr2/papers/TOP-18-005/trunk/TOP-18-005.tex $
\RCS$Id: TOP-18-005.tex 494881 2019-05-24 15:47:12Z otoldaie $

\newlength\cmsTabSkip\setlength{\cmsTabSkip}{1ex}
\providecommand{\cmsTable}[1]{\resizebox{\textwidth}{!}{#1}}
\providecommand{\NA}{\ensuremath{\text{---}}}

\cmsNoteHeader{TOP-18-005}
\title{Measurement of the top quark pair production cross section in dilepton final states containing one $\PGt$ lepton in $\Pp\Pp$ collisions at $\sqrt{s}=13\TeV$}

\date{\today}

\abstract{
The cross section of top quark pair production is measured in the $\ttbar\to (\ell\nu_{\ell})(\tauh\nu_{\PGt})\PQb\PAQb$ final state,
   where $\tauh$ refers to the hadronic decays of the $\PGt$ lepton, and $\ell$ is either an electron or a muon.
   The data sample corresponds to an integrated luminosity of 35.9\fbinv
   collected in proton-proton collisions at $\sqrt{s}=13\TeV$ with the CMS detector.
   The measured cross section is $\sigma_{\ttbar} = 781 \pm  7\stat \pm 62\syst \pm 20\lum\unit{pb}$, and
   the ratio of the partial width $\Gamma(\PQt\to\PGt\nu_{\PGt}\PQb)$ to the total decay width of the top quark is measured to be
   $0.1050 \pm 0.0009\stat \pm 0.0071\syst$.
   This is the first measurement of the \ttbar production cross section in proton-proton collisions at $\sqrt{s}=13\TeV$ that explicitly includes $\PGt$ leptons.
   The ratio of the cross sections in the $\ell\tauh$ and $\ell\ell$ final states
   yields a value $R_{\ell\tauh/\ell\ell}=0.973 \pm 0.009\stat \pm 0.066\syst$, consistent with lepton universality.
   }

\hypersetup{
pdfauthor={CMS Collaboration},
pdftitle={Measurement of the top quark pair production cross section in dilepton final states containing one tau lepton in pp collisions at 13 TeV},
pdfsubject={CMS},
pdfkeywords={CMS, physics, software, computing}}

\maketitle

\section{Introduction}

In proton-proton ($\Pp\Pp$) collisions at the CERN LHC, top quarks are produced mainly in pairs ($\ttbar$)
and subsequently decay to \PQb quarks and \PW bosons: $\Pp\Pp \to\ttbar\to \PWp\PQb\PWm\PAQb$.
The decay modes of the two \PW bosons determine the observed event signature.
The dilepton decay channel denotes the case where both \PW bosons
decay leptonically.
In this paper, we consider the process $\ttbar\to (\ell\PGn_\ell) (\PGt \PGnGt) \bbbar$,
where one \PW boson decays into
$\ell\PGn_\ell$ where $\ell$ is either an electron ($\Pe$) or a muon ($\PGm$), and the other into a tau lepton and a neutrino ($\PGt \PGnGt$).
The expected fraction of events
in this final state corresponds to $\approx$4/81 ($\approx$5\%)  of all \ttbar decays,
\ie equivalent to the fraction of all light dilepton channels ($\Pe\Pe$, $\PGm\PGm$, $\Pe\PGm$).

Recent checks of lepton flavour universality violation~\cite{Aaij:2017deq,Aaij:2019wad,Lees:2012xj,Lees:2013uzd,Bifani:2018zmi,Huschle:2015rga,Sato:2016svk,Hirose:2016wfn}
sparked a renewed interest towards measurements involving $\PGt$ leptons, owing to a potential disagreement with standard model (SM) predictions.
The $\PQt\to (\PGt\PGnGt) \PQb$ decay
exclusively involves third-generation leptons and quarks which, owing to their large masses, may be particularly sensitive to beyond SM contributions.
For example, the existence of a charged Higgs boson~\cite{Djouadi:2005gj,Branco:2011iw,Sirunyan:2019hkq,Aaboud:2018gjj}
may give rise to anomalous $\PGt$ lepton production that could be observed in this decay channel.

This is the first measurement of the \ttbar production cross section in $\Pp\Pp$ collisions at $\sqrt{s}=13\TeV$ that explicitly includes $\PGt$ leptons.
The data sample was collected in 2016 with the CMS detector at the LHC and corresponds to an integrated luminosity of 35.9\fbinv.
The $\PGt$ lepton is identified through its visible decay products, either hadrons ($\tauh$) or leptons ($\PGt_\ell$), with the corresponding branching fractions
$\mathcal{B}(\tauh \to \text{hadrons} + \PGnGt) \approx 65\%$ and
$\mathcal{B}(\PGt_\ell \to \ell~\PGn_{\ell}\PGnGt) \approx 35\%$.
In the first case, the $\tauh$ decays into a narrow jet with a distinct signature, whereas
the leptonic decays are difficult to distinguish from prompt electron or muon production.
In this measurement, the signal includes only $\PGt$ leptons that decay hadronically, and
$\ell$ does not include leptons from $\PGt$ decays.
The dominant background contribution comes from events where a jet is misidentified as a $\tauh$, mostly from
\ttbar lepton+jets events,
\ie $\ttbar\to (\ell\PGn_{\ell})(\PQq\PAQq')\bbbar$.
The cross section is measured by performing a profile likelihood ratio (PLR) fit~\cite{Cowan:2010js} to the transverse mass of the system
containing the lepton ($\Pe$ or $\PGm$) and the missing transverse momentum,
in two kinematic categories of the selected events for each of the $\Pe\tauh$ and $\PGm\tauh$ final states.
The cross section is measured in the fiducial phase space of the detector and also extrapolated to the full phase space.
The ratio of the cross sections in the $\ell\PGt$ and light dilepton~\cite{Sirunyan:2018goh} final states $\sigma_{\ttbar}(\ell\PGt)/\sigma_{\ttbar}(\ell\ell)$,
and the ratio of the partial to the total decay width of the top quark $\Gamma(\PQt\to\PGt\nu_{\PGt}\PQb)/\Gamma_{\text{total}}$ are evaluated.

This paper is organized as follows:
the CMS detector layout is briefly described in Section~\ref{sec:detector};
details about the simulated event samples used in the data analysis are provided in Section~\ref{sec:simulation};
Section~\ref{sec:eventsel} covers the event reconstruction and the event selection;
the event categorization and the fit procedures are described in Section~\ref{sec:eventcats};
the background determination procedure is given in Section~\ref{sec:background};
the description of the systematic uncertainties is presented in Section~\ref{sec:systematics};
measurements of the cross sections, and the ratio of the partial to the total \ttbar decay width are discussed in Section~\ref{sec:xsec};
and the results are summarized in Section~\ref{sec:summary}.

\section{The CMS detector}
\label{sec:detector}

The central feature of the CMS apparatus is a superconducting solenoid of 6\unit{m} internal diameter, providing a magnetic field of 3.8\unit{T}.
Within the solenoid volume are a silicon pixel and strip tracker, a lead tungstate crystal electromagnetic calorimeter, and a brass and scintillator hadron calorimeter, each composed of a barrel and two endcap sections,
covering
$0 < \varphi < 2\pi$ in azimuth and $\abs{\eta}<2.5$ in pseudorapidity.
Forward calorimeters extend the pseudorapidity coverage provided by the barrel and endcap detectors. Muons are detected in gas-ionization chambers embedded in the steel flux-return yoke outside the solenoid.
The detector is nearly hermetic, providing reliable measurement of the momentum imbalance in the plane transverse to the beams.
A two-level trigger system~\cite{Khachatryan:2016bia} selects the most interesting $\Pp\Pp$ collision events for use in physics analysis.
A more detailed description of the CMS detector, together with a definition of the coordinate system used and the relevant kinematic variables,
can be found in Ref.~\cite{Chatrchyan:2008zzk}.

\section{Event simulation}
\label{sec:simulation}

The analysis makes use of simulated samples of \ttbar events, as well as other processes that result in reconstructed $\PGt$ leptons in the final state.
These samples are used to design the event selection, to calculate the acceptance for \ttbar events, and to estimate most of the backgrounds in the analysis.

Signal \ttbar events are simulated with the \POWHEG event generator (v2)~\cite{Nason:2004rx,Frixione:2007vw,Alioli:2010xd,Campbell:2014kua,Frixione:2007nw}
at next-to-leading-order (NLO) accuracy in quantum chromodynamics (QCD).
The parton showers are modelled using
\PYTHIA (v8.2)~\cite{Sjostrand:2014zea}
with the CUETP8M2T4 underlying event (UE) tune~\cite{Skands:2014pea}.
The background samples used in the measurement of the cross section are simulated with \POWHEG and \MGvATNLO (v2.2.2)~\cite{Alwall:2014hca}.
The
\MGvATNLO
generator with MLM matching~\cite{Alwall:2007fs} is used for the simulation of \PW boson production in association with jets ({\PW}+jets),
and Drell--Yan (DY) production in association with jets at leading-order (LO) accuracy.
Here, only the leptonic decays of DY events and \PW bosons are simulated, and up to four additional jets are included.
The diboson processes are produced with NLO accuracy: $\PW\PW$ with \POWHEG, $\PW\cPZ$ and $\cPZ\cPZ$ with \MGvATNLO with FxFx matching~\cite{Frederix:2012ps}.
The \POWHEG generator is used for the simulation of $t$-channel single top quark production and
single top quark production associated with a \PW boson ($\PQt\PW$)~\cite{Re:2010bp,Alioli:2009je}.
The single top quark $s$-channel sample is produced with \MGvATNLO at NLO accuracy with FxFx matching scheme.
The simulated events are produced with a top quark mass of $m_{\PQt} = 172.5\GeV$.
The generated events are subsequently processed with
\PYTHIA
using the underlying event tune CUETP8M1
to provide the showering of the partons, and to perform the matching of the soft radiation
with the contributions from direct emissions included in the matrix-element (ME) calculations.
The default
parton distribution functions (PDFs)
are
the NNPDF3.0~\cite{Ball:2014uwa}.
The $\PGt$ decays are simulated with \PYTHIA,
which correctly accounts for the $\PGt$
lepton polarization in describing the kinematic properties of the decay.
The CMS detector response is simulated with \GEANTfour~\cite{Agostinelli:2002hh}.
Additional pp interactions in the same or nearby bunch crossings (pileup, PU) are superimposed on the hard collision.
Simulated events are reweighted to match the distribution of the number of pileup collisions per event in data.
This distribution is derived from the instantaneous luminosity and the inelastic cross section~\cite{Sirunyan:2018nqx}.

The next-to-next-to-leading-order (NNLO) expected SM \ttbar pair production cross section
of $832^{+20}_{-29}\,\text{(scale)}\pm35\,\mathrm{(PDF+}\alpS)\unit{pb}$~\cite{Czakon:2011xx} ($m_{\PQt}=172.5\GeV$)
is used for the normalization of the number of \ttbar events in the simulation.
The first uncertainty includes the uncertainties in the factorization
and renormalization
scales,
while the second is  associated with possible choices of PDFs and the value of the strong coupling constant ($\alpS$).
The proton structure is described by the CT14~(NNLO) PDF set with the corresponding PDF and $\alpS$ uncertainties~\cite{Dulat:2015mca}.
The {\PW}+jets and DY+jets backgrounds are normalized to their NNLO cross sections calculated with \textsc{fewz} (v3.1)~\cite{Li:2012wna}.
The $t$-channel and the $s$-channel single top quark production are normalized to the NLO calculations obtained from \textsc{Hathor} (v2.1)~\cite{Aliev:2010zk,Kant:2014oha}.
The production of $\PQt\PW$ is normalized to the NNLO calculation~\cite{Kidonakis:2010ux,Kidonakis:2012rm}.
Finally, the production of diboson pairs is normalized to the NLO cross section prediction calculated with \textsc{mcfm}~\cite{Campbell:2010ff,Campbell:2011bn} (v7.0).

\section{Event reconstruction and selection}
\label{sec:eventsel}

The signal event topology is defined by the presence of two $\PQb$ quark jets from the top quark decays,
one \PW\ boson decaying leptonically into $\Pe\PGn$ or $\PGm\PGn$, and a second \PW\ boson decaying into $\tauh\nu$.
In each event, all objects are reconstructed with a particle-flow (PF) algorithm~\cite{Sirunyan:2017ulk}.
This algorithm combines the information from all subdetectors to identify and
reconstruct all types of particles in the event, namely charged and neutral hadrons, photons, muons, and electrons, together referred to as PF objects.
These objects are used to construct
a variety of higher-level objects and observables, including jets and missing transverse momentum ($\ptvecmiss$), which is
the negative vector sum of transverse momenta of all reconstructed PF objects.
Parameters of jets and the tracks associated with jets provide input variables for \PQb tagging discriminators.
The reconstructed vertex with the largest value of summed physics-object $\pt^2$ is taken to be the primary $\Pp\Pp$ interaction vertex.
Jets are reconstructed by clustering PF objects
with the anti-\kt~\cite{Cacciari:2008gp}
jet algorithm with a distance parameter
$R = 0.4$.

Electron or muon candidates are required to originate from the primary vertex, pass quality selection criteria, and be isolated relative to other activity in the event.
The relative isolation is based on PF objects within a cone of $\Delta R = \sqrt{\smash[b]{(\Delta\eta)^2+(\Delta\varphi)^2}} = 0.4$ around the electron or muon track,
and defined as
$I_\text{rel} = (E_\text{ch} + E_\text{nh} + E_\text{ph} - 0.5 \, E^\text{PU}_\text{ch})/\pt$, where
$E_\text{ch}$ is the transverse energy deposited by charged hadrons from the primary vertex,
$E_\text{nh}$ and $E_\text{ph}$ are the respective transverse energies of the neutral hadrons and photons,
and
$0.5 \, E^\text{PU}_\text{ch}$ is the estimation of the contribution of neutral particles from pileup vertices,
calculated as half of the energy of the charged particles from pileup;
\PT is the electron or muon transverse momentum.
Electron candidates with $I_\text{rel} < 0.0588$ in the barrel or $I_\text{rel} < 0.0571$ in the endcaps are considered isolated.
The muon candidate is isolated if $I_\text{rel} < 0.15$ in either the barrel or the endcaps.
The lepton isolation requirements are used to suppress backgrounds from multijet production.
The charge misidentification probability for electrons and muons is less than 0.5\% and 0.1\%, respectively,
and is measured from \PZ boson decays and simulation~\cite{Khachatryan:2015hwa,CMS-DP-2017-004,Sirunyan:2018fpa}.

Hadronic $\PGt$ lepton decays are reconstructed with the hadron-plus-strips (HPS) algorithm~\cite{Sirunyan:2018pgf},
which starts from reconstructed jets.
In each jet, a charged hadron is combined with other nearby charged hadrons or photons to identify the decay modes.
The identification of $\Pgpz$ mesons is enhanced by clustering electrons and photons in
``strips" along the track bending direction to take into account possible broadening of calorimeter signatures by early showering photons.
The $\tauh$ candidates are selected from the following combinations of charged hadrons and strips that correspond to the $\PGt$ decay modes:
single hadron, hadron plus a strip, hadron plus two strips, and three hadrons.
A multivariate analysis of these HPS $\tauh$ candidates is used to reduce the contamination from quark and gluon jets.
A boosted decision tree is trained using a sample of DY events with $\tauh$ decays as signal and a sample of QCD multijet events as background, both from simulation.
Input variables include the multiplicity and the transverse momenta of electron and photon candidates in the vicinity of the $\tauh$,
the kinematic properties of hadrons and strips,
and the $\tauh$ lifetime information, such as the impact parameter of the leading track and the significance of the length of flight to the secondary vertex of the $\tauh$ candidates with three charged hadrons.
Additional requirements are applied to discriminate genuine $\tauh$ leptons from prompt electrons and muons.
The $\tauh$ charge is taken as the sum of the charges of the corresponding charged hadrons.
The misidentification probability for the charge is less than 1\% and it is estimated from $\PZ\to \PGt\PGt\to\PGm\tauh$ data events with same-charge $\PGm$ and $\tauh$.
The $\tauh$ identification efficiency of this algorithm is estimated to be approximately 60\%
for $\pt>20\GeV$, and it is measured in a sample enriched in $\PZ\to\Pgt\Pgt\to\Pgm\tauh$ data events
with a ``tag-and-probe'' technique~\cite{Sirunyan:2018pgf}.
The corresponding probability for generic hadronic jets to be misidentified as $\tauh$ is less than 1\%~\cite{Sirunyan:2018pgf}.

For the $\Pe\tauh$ ($\PGm\tauh$) final state,
data are collected with a trigger requiring
at least one isolated electron (muon) with a threshold of $\pt>27$ (24)\GeV.

\begin{figure}[htp]
\centering
\includegraphics[width=0.48\textwidth]{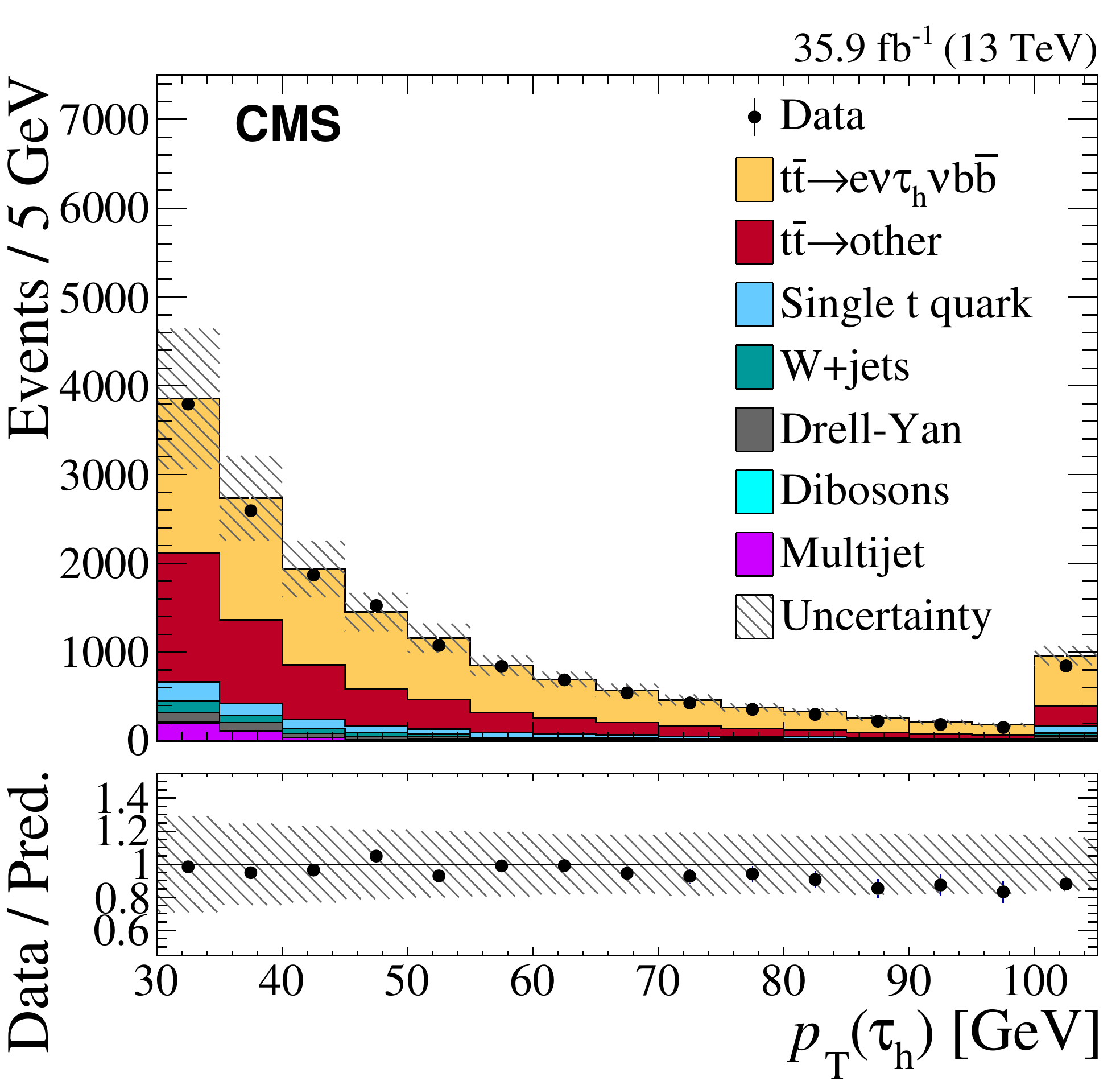}
\includegraphics[width=0.48\textwidth]{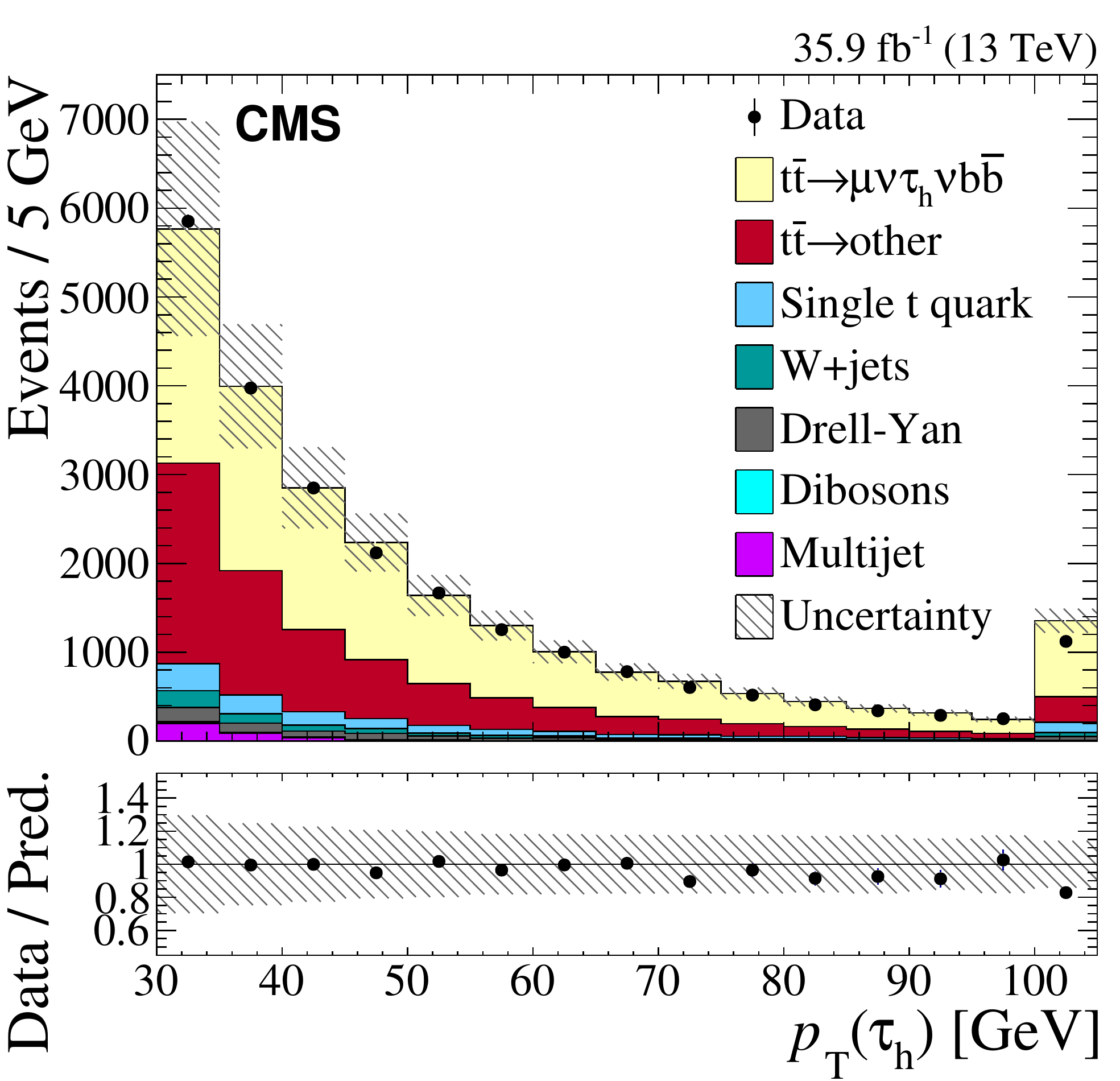}
\caption{
The $\tauh$ $\pt$ distributions
for events of the $\Pe\tauh$ (left) and $\PGm\tauh$ (right) final states
observed prior to fitting.
Distributions obtained from data (filled circles) are compared with simulation (shaded histograms).
The last bin includes overflow events.
The simulated contributions are normalized to the cross section values predicted in the SM.
The main processes are shown:
the signal,
the other \ttbar processes grouped together,
single top quark production,
{\PW}+jets,
DY processes,
diboson, and multijet production.
The ratio of the data to the total SM prediction is shown in the lower panel.
The hatched bands indicate the systematic uncertainties and the statistical uncertainties of all simulated samples.
Statistical uncertainties on the data points are not visible because of the scale of the figure.
}
\label{fig:taupt}
\end{figure}

Events are selected by requiring one isolated electron (muon)
with transverse momentum $\pt>30$ (26)\GeV and $\abs{\eta}<2.4$,
at least two jets with $\pt>30\GeV$ and $\abs{\eta}<2.5$,
and exactly one $\tauh$ candidate with $\pt> 30\GeV$ and $\abs{\eta}<2.4$.
The $\tauh$ candidate and the selected lepton are required to have opposite electric charges (OC).
Electrons or muons are required to be separated from any jet and from the $\tauh$ candidate in the
$\eta$-$\varphi$ plane by a distance $\Delta R > 0.4$.
Events with any additional loosely isolated electron (muon) of $\pt>15$ (10)\GeV are rejected.
An electron is considered loosely isolated if $I_\text{rel} < 0.0994$ in the barrel or $I_\text{rel} < 0.107$ in the endcaps.
A muon is loosely isolated if $I_\text{rel} < 0.25$ in either the barrel or the endcaps.
At least one jet is required to be identified as originating from \PQb\ quark hadronization (``\PQb\-tagged'').
The \PQb tagging algorithm used (``CSVv2'' in Ref.~\cite{Sirunyan:2017ezt})
combines the information of displaced tracks and secondary vertices associated with the jet in a multivariate technique.
The working point selected provides a \PQb tagging efficiency of about 66\% with a corresponding light-flavour misidentification rate of 1\%.
The selected events exhibit good agreement between the observed data and the expectation,
as shown in Fig.~\ref{fig:taupt} for the $\pt$ distribution of the $\tauh$ candidate.
The dominant background contribution comes from other \ttbar decays, mostly from lepton+jets final states where a jet is misidentified as a $\tauh$ candidate.

\section{Event categories and fit procedure}
\label{sec:eventcats}

\begin{figure}[htp]
\centering
\includegraphics[width=0.48\textwidth]{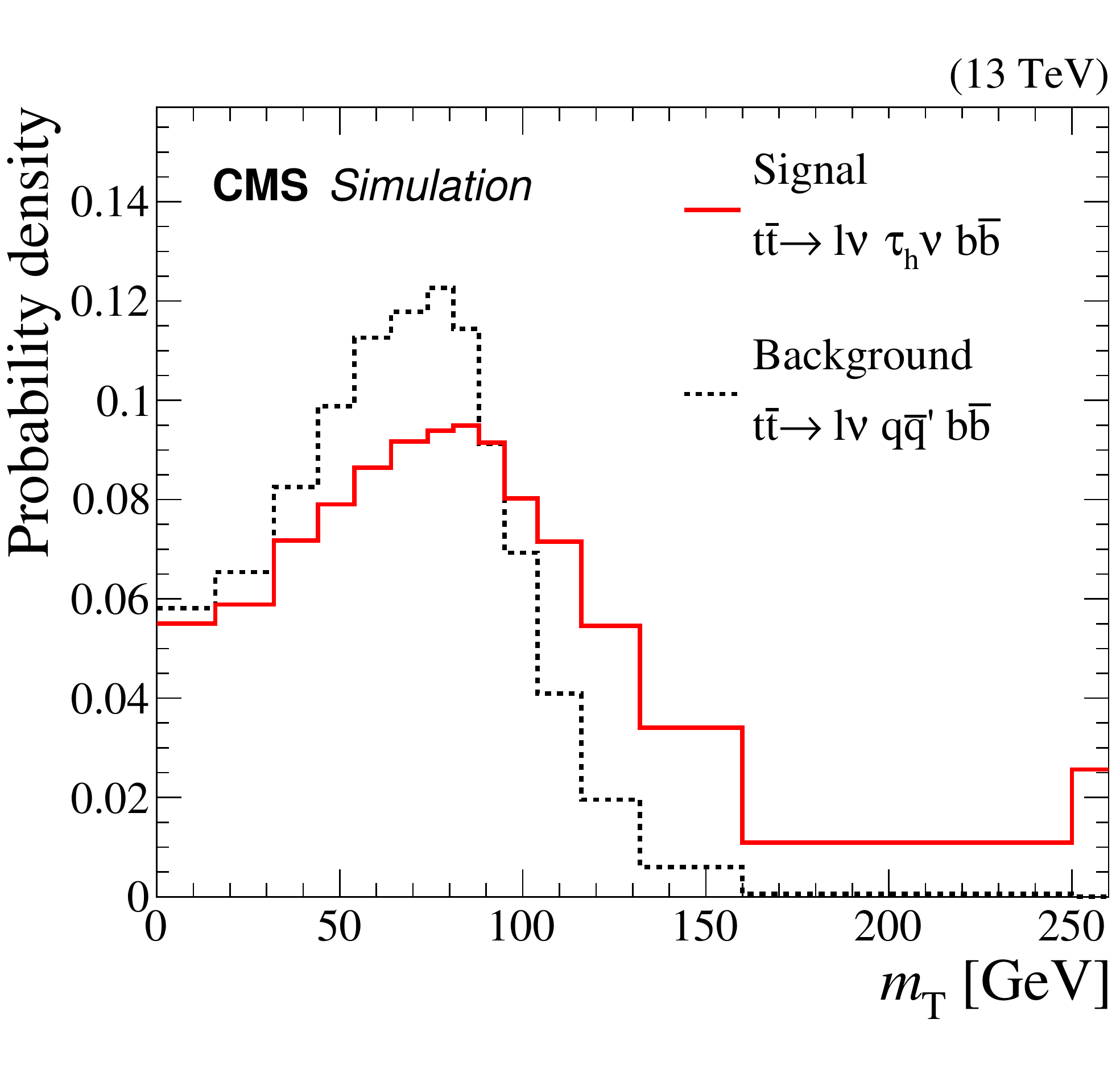}     \hfill
\includegraphics[width=0.48\textwidth]{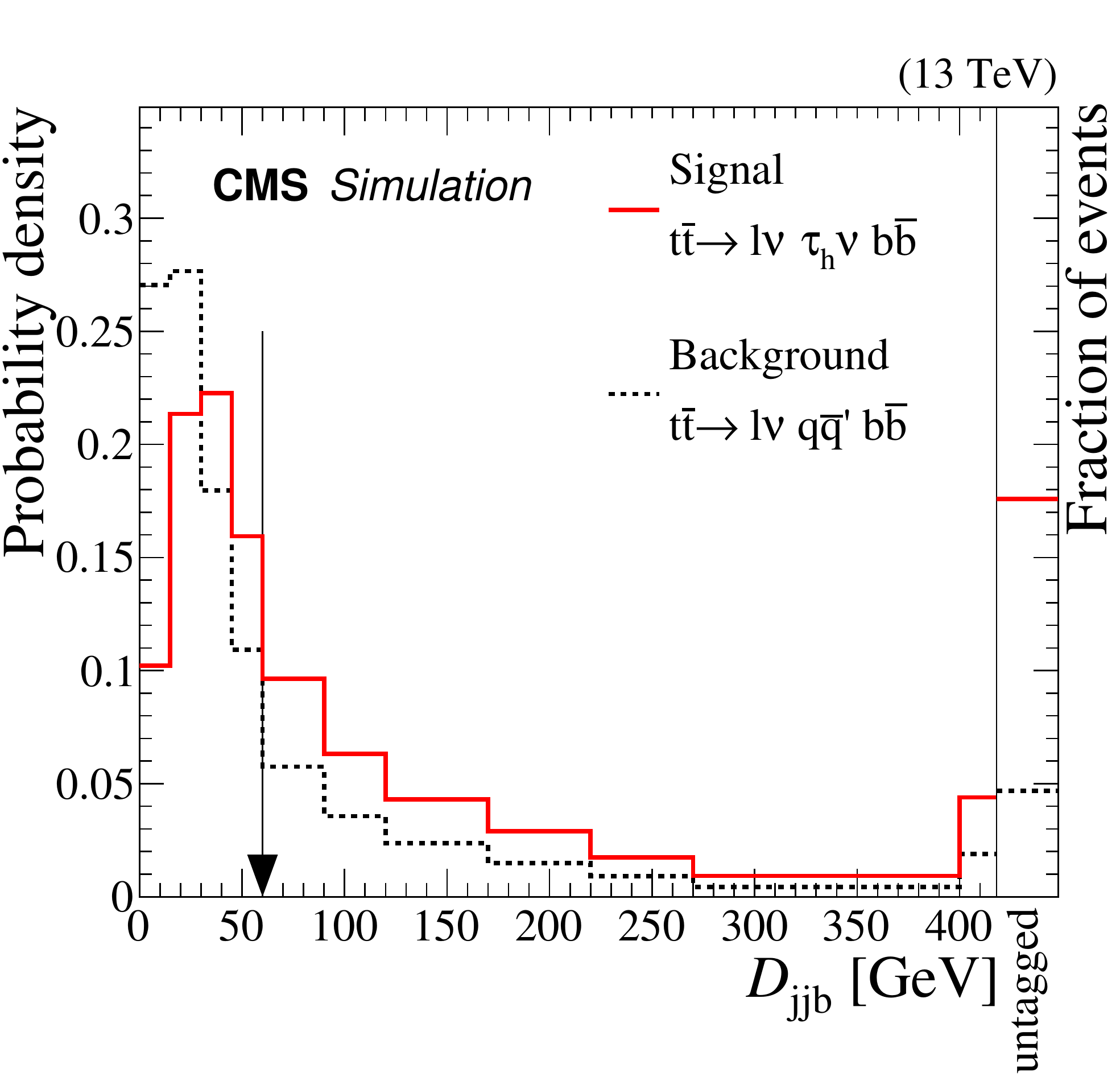}
\caption{
Comparison of the signal ($\ttbar \to \ell\nu_{\ell} \tauh\nu_{\PGt} \bbbar$)
and the main background of misidentified $\tauh$ ($\ttbar \to \ell\nu_{\ell} \PQq\PAQq' \bbbar$)
in the shapes of the normalized distributions of the transverse mass $\mT$ between the lepton and \ptmiss (left),
and the $D^\text{min}_{\mathrm{jjb}}$ parameter (see text) of the event categories (right).
In the $\mT$ distribution, the signal may extend beyond the \PW boson mass endpoint because of the two-neutrino final state,
whereas the background process cannot.
The last bin in both distributions includes overflow events.
In the $D^\text{min}_{\mathrm{jjb}}$ distribution,
the downward arrow points at the threshold of the cut used ($D^\text{min}_{\mathrm{jjb}}>$60\GeV),
and the panel on the right shows the fraction of events in the ``signal-like'' category where there is only one untagged jet,
which amounts to approximately 5\% of all background events and 17\% of all signal events.
}
\label{fig:mt_shapes}
\end{figure}

The \ttbar production cross section is extracted from a
PLR fit of the binned distribution
of the transverse mass of the lepton and \ptmiss in two kinematic event categories, for each of the $\Pe\tauh$ and $\PGm\tauh$ final states.
The transverse mass is defined as $\mT=\sqrt{\smash[b]{2 \abs{\ptvec^{\ell}} \abs{\ptvecmiss} (1-\cos\Delta\varphi)}}$,
where $\Delta\varphi$ is the azimuthal angle difference between the lepton transverse momentum vector, $\ptvec^{\ell}$, and $\ptvecmiss$.
The $\mT$ distribution
provides separation between signal and background processes (as shown in Fig.~\ref{fig:mt_shapes})
and does not significantly depend on \pt and $\eta$ of the \PGt candidate, or other jet characteristics in the kinematic ranges of this study.
Two event categories are defined according to the kinematic properties of jets in the event.
In order to discriminate against the main background of misidentified $\tauh$ from the \ttbar lepton+jets process,
the constraints from top quark and \PW boson masses in the decay $\PQt\to \PQb \PW \to \PQb (\PQq\PAQq')$ are used.
Jet triplets are constructed for each combination of one \PQb-tagged jet and two untagged jets,
chosen from all jets in the event, including the $\tauh$ candidate.
The distance parameter for each triplet is calculated as
$D_{\mathrm{jjb}} = \sqrt{\smash[b]{(m_{\PW} - m_{\mathrm{jj}})^2 + (m_{\PQt} - m_{\mathrm{jjb}})^2}}$,
where $m_{\PQt} = 172.5\GeV$ and $m_{\PW} = 80.385\GeV$ are, respectively, the masses of the top quark and of the \PW boson~\cite{PDG2018},
$m_{\mathrm{jj}}$ is the invariant mass of the two untagged jets,
and $m_{\mathrm{jjb}}$ is the invariant mass of the jet triplet.
The event is assigned to the ``signal-like'' category
if there is only one untagged jet,
or if the minimum parameter value $D^\text{min}_{\mathrm{jjb}}$ is larger than 60\GeV.
Otherwise, it is assigned to the ``background-like'' event category.
The threshold of 60\GeV provides an optimal separation of signal and background event categories, together with a maximization of the yields in each of the two categories
in order to reduce the statistical uncertainties.
In the fit, the two event categories provide an additional constraint on the background processes
independent from the details of the $\mT$ distribution.

The cross section is derived from the signal strength measured in the fit, \ie its ratio to the value expected in the SM.
It is estimated for both event categories, in each of the $\Pe\tauh$ and $\PGm\tauh$ final states.
The expected number of events in a given bin of the $\mT$ distribution
is parametrized as a function of signal strength and nuisance parameters.
The nuisance parameters encode the effects of systematic uncertainties.
The signal strength is a free parameter in the fit.
The fitted variables do not significantly depend on the kinematic properties of the $\PGt$ lepton in the specific \ttbar signal model considered here,
\ie $\ttbar\to (\ell\nu_{\ell})(\tauh\nu_{\PGt})\bbbar$.
The likelihood function is defined as a product of Poisson distributions of the expected number of events in bins of the $\mT$ distribution and nuisance constraints.
Based on the likelihood function, the PLR test statistic is defined as
the ratio between the
maximum of the likelihood for a given value of signal strength
and the global maximum of the likelihood function.
The effect of the systematic uncertainties on the signal strength is determined with this approach.

\section{Background estimate}
\label{sec:background}

The main background contribution comes from events with one lepton, significant \ptmiss, and three or more jets,
dominated by the lepton+jets \ttbar process,
where one of the jets is falsely identified as a $\tauh$.
Misidentified $\tauh$ candidates also come from multijet and {\PW}+jet background processes.
There is a small contribution from processes with genuine hadronic $\tauh$:
$\PQt\PW$ single top quark production, $\tau_{\ell}\tauh$ from DY decays,
$\ttbar\to \tau_{\ell} \tauh \bbbar$, and diboson processes.
All processes, except multijet, are estimated from simulation after applying appropriate corrections.
The pileup,
trigger efficiencies, lepton identification, jet energy corrections, and \PQb tagging efficiencies in the simulation
are corrected with scale factors measured in separate publications~\cite{Sirunyan:2018fpa, Khachatryan:2015hwa, PILEUP},
as described in Section~\ref{sec:systematics}.

The $\tauh$ misidentification contribution is determined by constraining the falsely identified $\tauh$
in the overall fit to the data in the $\mT$ distribution.
In the fit, the event yields of the background processes with a misidentified $\tauh$ are determined
by adjusting the normalization of the shapes of the $\mT$ distributions.
The normalization factors are introduced as nuisance parameters with constraints determined from studies in other processes~\cite{Sirunyan:2018pgf}.
The corresponding uncertainties are discussed in Section~\ref{sec:systematics}.

The background from the multijet processes is determined from data as it provides a more accurate description with a smaller statistical uncertainty.
The shape of the $\mT$ distribution is obtained from a sample of events containing lepton and $\tauh$ candidates of the same charge (SC).
It is estimated by subtracting from the data all other processes taken from simulation, including the fully hadronic final states in \ttbar, single top quark, and dibosons.
The $\mT$ shapes for SC and OC events are the same within the uncertainties in a control region with a relaxed $\tauh$ identification requirement, and in agreement with the simulation.
The normalization is corrected by multiplying the SC $\mT$ distribution by the OC-to-SC ratio, $f_{\mathrm{OC/SC}}$,
as determined in a control region from events with a relaxed $\tauh$ identification and an inverted lepton isolation requirement, where the multijet contribution is dominant.
All other event selection requirements remain the same as in the main selection.
The ratio is measured to be $f_{\mathrm{OC/SC}} = 1.05 \pm 0.05\,\text{(stat + syst)}$,
in agreement with simulation.
As one of the processes with misidentified $\tauh$, the normalization of the multijet contribution is varied in the fit as a separate nuisance parameter,
as described in Section~\ref{sec:systematics}.

\section{Systematic uncertainties}
\label{sec:systematics}

The main sources of systematic uncertainty are from
$\tauh$ identification and misidentification, \PQb tagging, estimation of pileup in the $\Pp\Pp$ collisions, jet energy scale (JES), and jet energy resolution (JER).
Other sources of uncertainty are from lepton identification, trigger efficiency,
and the calibration of the integrated luminosity.
Theoretical uncertainties are also included in the event simulation.
Uncertainties are applied in a coherent way to signal and background processes.
The corresponding corrections and their uncertainties
are measured in dedicated studies,
which are described below.

The uncertainty in the efficiency of $\tauh$ identification is 5\% for all $\tauh$ with $\pt > 20\GeV$ and is applied to all processes with a genuine $\tauh$.
It is measured with a tag-and-probe
technique in samples enriched in $\PZ\to\tau_{\ell}\tauh$ events~\cite{Sirunyan:2018pgf}.
The \tauh charge confusion probability, estimated to be less than 1\%, is considered a part of the \tauh identification efficiency uncertainty.
The correction to the reconstructed energy of the $\tauh$ jet ($\PGt$ energy scale) and the corresponding uncertainty
is estimated in a fit of the data in distributions sensitive to the $\PGt$ energy, such as the $\tauh$ visible mass~\cite{Sirunyan:2018pgf}.
The dominant background contribution arises from processes where a jet is misidentified as $\tauh$,
mainly lepton+jets \ttbar, {\PW}+jets, and multijet production.
The $\tauh$ misidentification probability and its uncertainty in these processes are directly measured in the fit.
The misidentification probability is varied within $\pm50\%$ of the expected values in all processes with a jet falsely identified as the $\tauh$ candidate.
The variation covers the differences between expected and observed misidentification probabilities
and the possible dependence on other kinematic properties of the $\tauh$ candidate~\cite{Sirunyan:2018pgf}.
The misidentification probability is significantly constrained in the fit
and is not the dominant source of the uncertainty in the final result.

The uncertainties related to \PQb tagging (mistagging) efficiencies are estimated from
a variety of control samples enriched in \PQb~quarks (\PQc and light-flavour quarks)~\cite{Sirunyan:2017ezt};
the data-to-simulation scale factors for \PQb, \PQc, and light-flavour jets are applied to the simulation and the corresponding uncertainties are included in the fit.

The uncertainties in the JES, JER, and \ptmiss scales are estimated according to the prescription described in Ref.~\cite{JET}.
The uncertainty in the JES is evaluated as a function of jet \pt and $\eta$.
The JES and JER uncertainties are propagated to the \ptmiss scale.

The lepton trigger, identification, and isolation efficiencies are measured in data and simulation
with a tag-and-probe method
in $\PZ\to \ell^{+}\ell^{-}$ events~\cite{Khachatryan:2015hwa,CMS-DP-2017-004,Sirunyan:2018fpa}.
The simulated events are corrected with the corresponding data-to-simulation scale factors.
The uncertainties in the scale factors are included as systematic uncertainties in the measurement.

The uncertainty in the integrated luminosity is estimated to be 2.5\%~\cite{CMS:2017sdi}.

The pileup distribution is estimated from the measured luminosity in each bunch crossing multiplied by the average total inelastic cross section.
It is used to model the pileup in simulation with an uncertainty obtained by varying the inelastic pp cross section extracted from a control region by its uncertainty of $\pm4.6\%$~\cite{PILEUP}.

The measurement includes the uncertainty in the modelling of the \PQb quark fragmentation,
which covers $\Pep\Pem$ data~\cite{Abbiendi:2002vt,Heister:2001jg,Abe:2002iq,Abdallah2011} at the \PZ
pole with the Bowler--Lund~\cite{Bowler1981}
and Peterson~\cite{PhysRevD.27.105} parametrizations, and
the uncertainties in the semileptonic \PQb-flavoured hadron branching fractions according to their 
measured values~\cite{PDG2018}.
An uncertainty in the modelling of the \pt distribution of the top quark in \ttbar processes
is included to cover the difference between the predicted and observed spectra~\cite{Sirunyan:2017mzl,Khachatryan:2016mnb,Khachatryan:2015oqa,Khachatryan:2015fwh}.
The fit is sensitive to the top quark \pt as it affects the shape of the $\mT$ distribution.
The top quark \pt variation also covers the slight trend of the $\tauh$ \pt distribution.

The cross section is measured by the fit in the fiducial phase space of the detector.
The fiducial cross section is extrapolated to the full phase space by correcting for the acceptance of the \ttbar signal process.
The fit and the acceptance include the following modelling uncertainties:
the renormalization and factorization scales, and PDFs including $\alpS$.
The uncertainty in the PDF is estimated by using the CT14~(NNLO) set as alternative PDFs.
The renormalization and factorization scales in the ME calculations are varied independently by factors of 0.5 and 2.0 from their nominal values,
and the envelope of the variations is included in the measurement.
The scale is varied by factors of 0.5 and 2.0 in the parton shower (PS) simulation of final-state and initial-state radiation, FSR and ISR.
The $h_{\text{damp}}$ parameter regulating the real emissions in \POWHEG (ME-PS matching) is
varied from its central value of $1.58\,m_{\PQt}$ using samples with $h_{\text{damp}}$ set to $0.99\,m_{\PQt}$ and $2.24\,m_{\PQt}$ ($m_{\PQt} = 172.5\GeV$),
as obtained from tuning this parameter to \ttbar data at $\sqrt{s} = 8\TeV$~\cite{CMS-PAS-TOP-16-021}.
The underlying event tune is varied within its uncertainties~\cite{Skands:2014pea,CMS-PAS-TOP-16-021}.
The effect of these uncertainties on the final state objects is included in the fit in the fiducial phase space
by adding the corresponding systematic variations normalized to the nominal acceptance.
Therefore, the measurement in the fiducial phase space is performed with the nominal acceptance and its uncertainties
are only included in the extrapolation to the full phase space.
The uncertainties in the fit are not correlated with the acceptance uncertainty in the extrapolation to the full phase space.

The theoretical uncertainties are implemented by reweighting the simulated events with corresponding scale factors.
The differences between weighted and unweighted distributions are taken as the uncertainties in the modelling.
Separate data sets with varied parameters are used for determining FSR, ISR, ME-PS matching, and underlying event uncertainties.

The impact of the systematic uncertainties on the measurement is given in Table~\ref{tab:systsall}.

\section{Results}
\label{sec:xsec}

\begin{figure}[htp]
\centering
\includegraphics[width=0.48\textwidth]{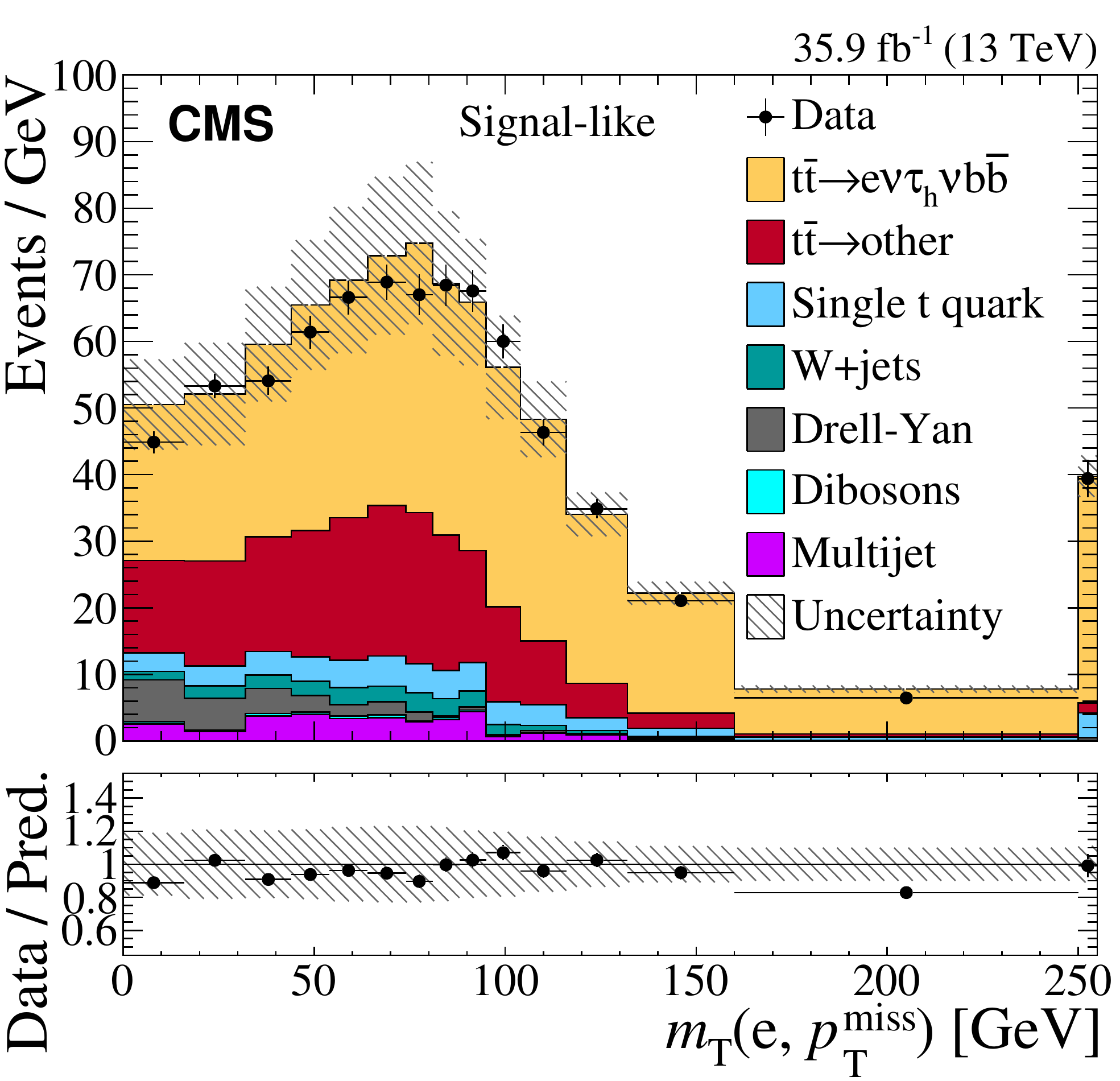} \hfill
\includegraphics[width=0.48\textwidth]{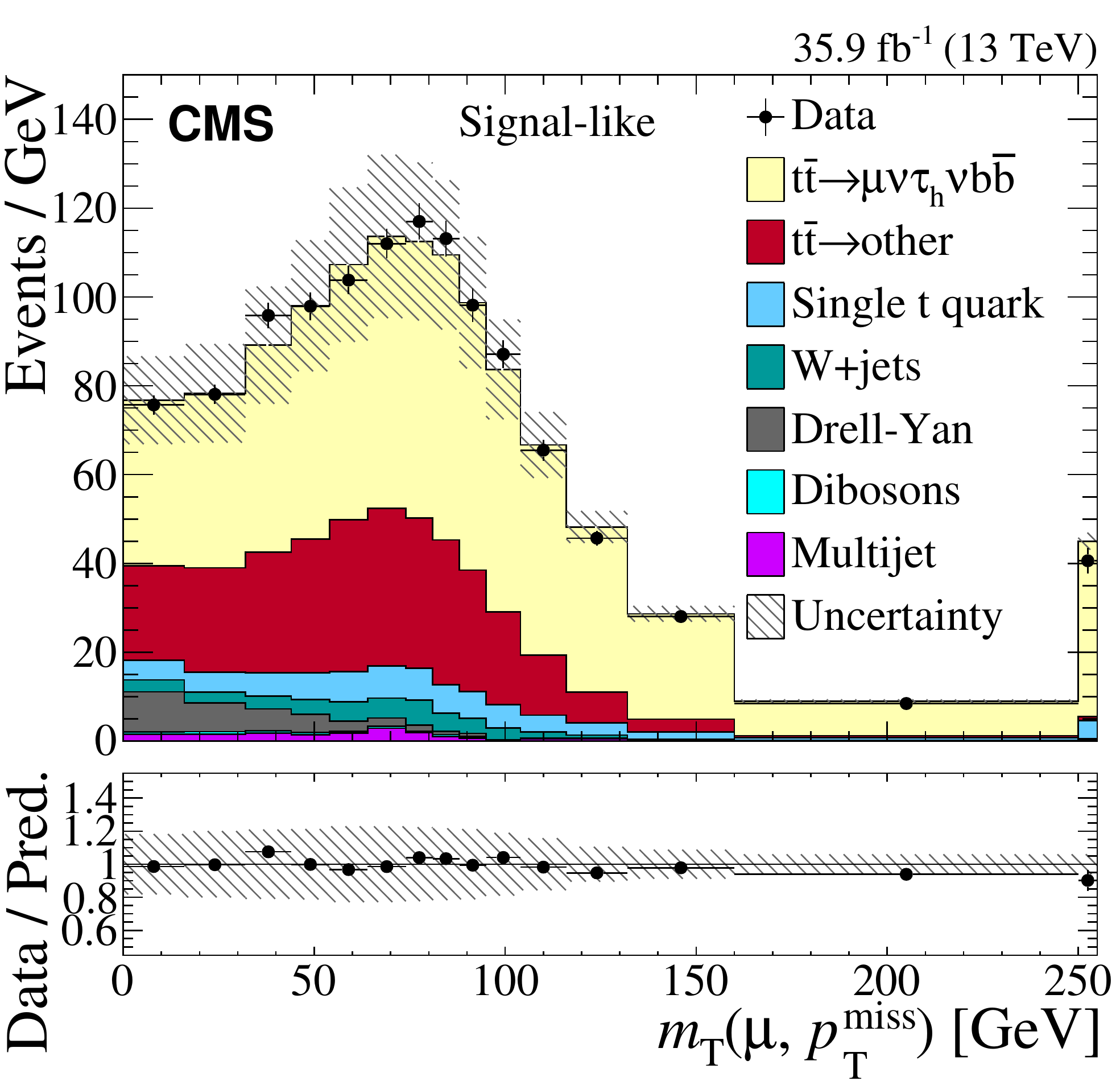} \\
\includegraphics[width=0.48\textwidth]{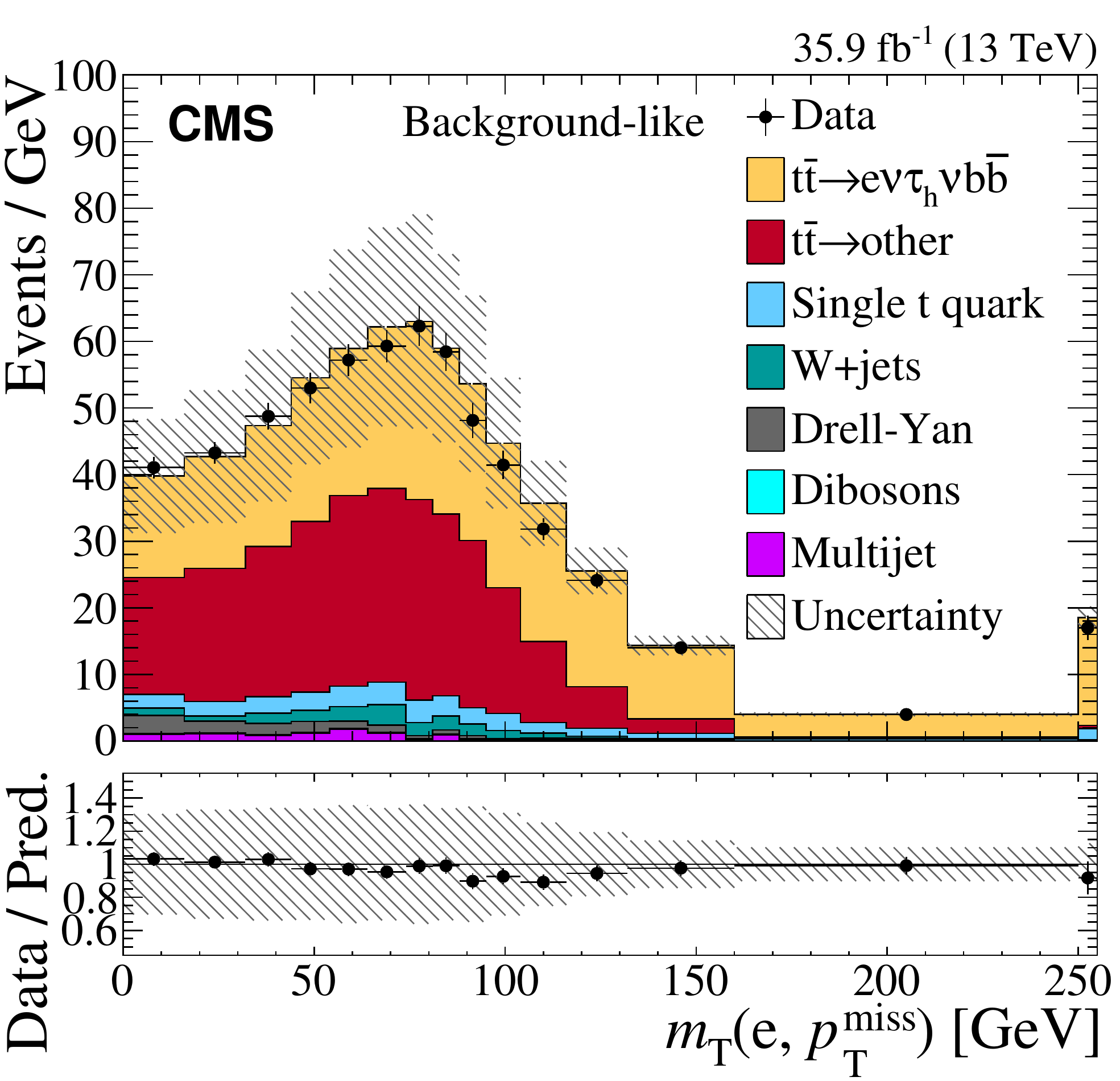} \hfill
\includegraphics[width=0.48\textwidth]{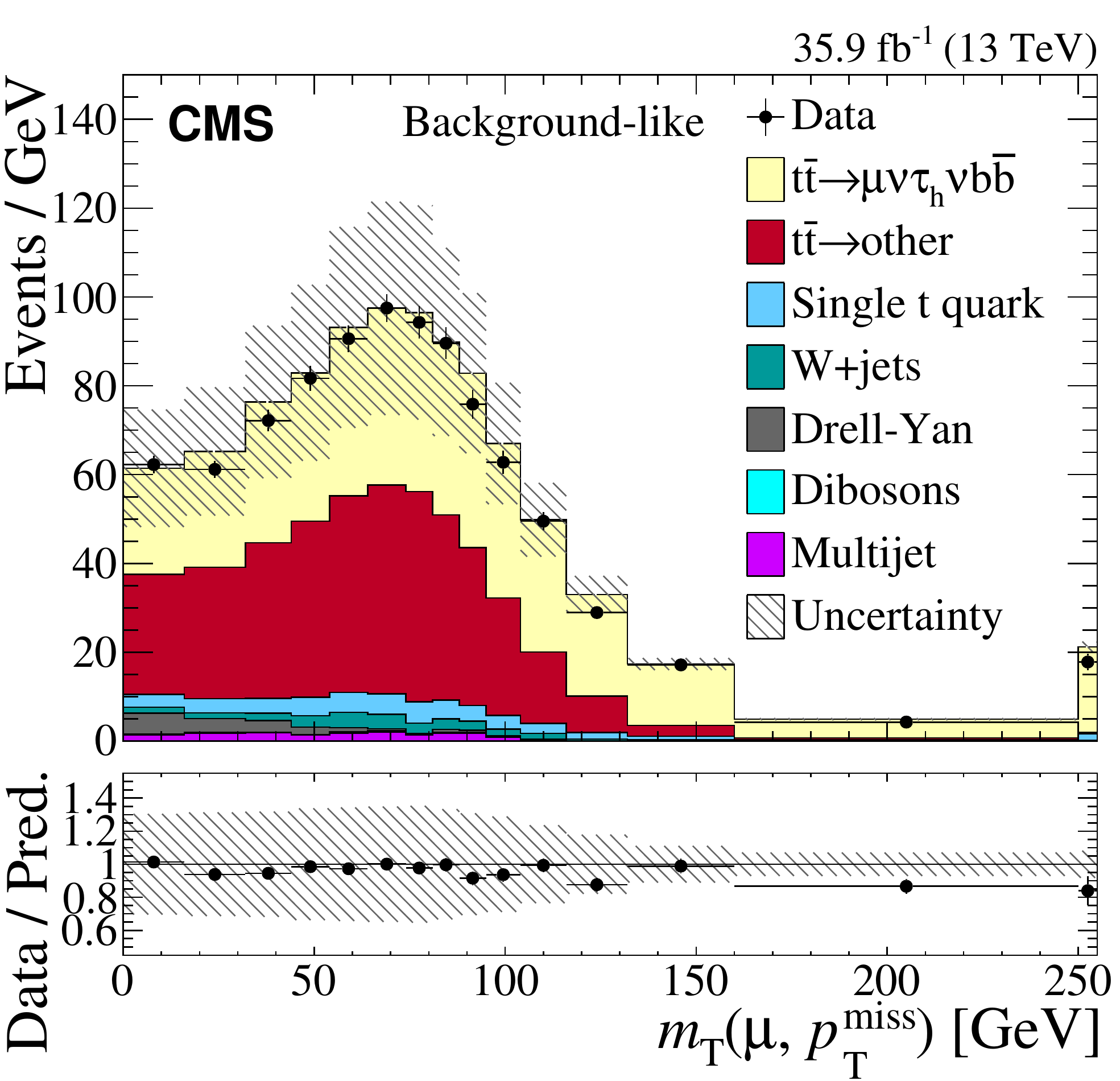}
\caption{
The transverse mass distributions between lepton ($\Pe$ or $\PGm$) and \ptmiss, $\mT$,
in the signal-like (upper) and background-like (lower) event categories
for the $\Pe\tauh$ (left) and $\PGm\tauh$ (right) final states
observed prior to fitting.
Distributions obtained from data (filled circles) are compared with simulation (shaded histograms).
The last bin includes overflow events.
The simulated contributions are normalized to the cross section values predicted in the SM.
The main processes are shown:
the signal,
the other \ttbar processes grouped together,
single top quark production,
{\PW}+jets,
DY processes,
diboson, and multijet production.
The ratio of the data to the total SM prediction is shown in the lower panel.
The vertical bars on the data points indicate the statistical uncertainties,
the hatched band indicates the systematic uncertainties and the statistical uncertainties in all simulated samples.
}
\label{fig:mtdistribution}
\end{figure}

The event yields expected from the signal and background processes,
as well as the observed event yields are summarized in Table~\ref{tab:SummaryEventYield},
for the signal-like and the background-like event categories (described in Section~\ref{sec:eventcats}) in each of the $\Pe\tauh$ and $\PGm\tauh$ final states.
The observed event yields in data show good agreement with the prediction.
The $\mT$ distributions
in the two categories of the selected events
are shown in Fig.~\ref{fig:mtdistribution}, for both the $\Pe\tauh$ and $\PGm\tauh$ final states.
A good shape agreement is observed between the data and the expected sum of signal and background distributions.

\begin{table}[htp]
\centering
\topcaption{Expected and observed event yields in the $\ell\tauh$ ($\ell=\Pe,\PGm$) final state for signal
and background processes for an integrated luminosity of 35.9\fbinv. Statistical and systematic uncertainties are shown.
The expected prefit contributions of all processes are presented separately for background-like and signal-like event categories.
The statistical uncertainties of the modelling are shown for the processes estimated from the simulation.
The multijet contribution and the corresponding statistical uncertainties are estimated using data, as described in Section~\ref{sec:background}.
}
\label{tab:SummaryEventYield}

\cmsTable{
\begin{tabular}{lcccc}
\hline
  & \multicolumn{4}{c}{Number of events ($\pm \text{stat} \pm \text{syst}$)}        \\
  & \multicolumn{2}{c}{$\Pe\tauh$}     &   \multicolumn{2}{c}{$\PGm\tauh$}   \\
\multicolumn{1}{c}{Process}  & Background-like &  Signal-like & Background-like & Signal-like \\
\hline
Signal & & & & \\
\hspace*{4mm}$\ttbar\to (\ell\nu_{\ell})(\tauh\nu_{\PGt})\bbbar$ & $3440\pm40\pm210$ & $5320\pm40\pm360$ &  $5140\pm40\pm 130$ & $7890\pm50 \pm 280$  \\ [\cmsTabSkip]
$\ttbar$ backgrounds & & & & \\
\hspace*{4mm}$\ttbar\to (\ell\nu_{\ell})(\PQq\PAQq')\bbbar$ & $2450\pm30\pm1210$ & $1610\pm20\pm830$ & $3670\pm40\pm1810$ & $2440\pm30\pm 1260$  \\
\hspace*{4mm}$\ttbar\to\text{other}$ & $390\pm10\pm70$ & $510\pm10\pm80$ & $580\pm10\pm110$ & $760\pm20\pm120$  \\ [\cmsTabSkip]
Other backgrounds & & & & \\
\hspace*{4mm}Single \PQt quark & $370\pm10\pm90$ & $540\pm10\pm100$ & $500\pm10\pm110$ & $790\pm10\pm150$  \\
\hspace*{4mm}Drell--Yan & $150\pm20\pm20$ & $310\pm20\pm20$ & $200\pm20\pm10$ & $410\pm30\pm40$ \\
 [\cmsTabSkip]
Total & $7090\pm80\pm1230$  & $8930\pm80\pm920$ & $10\,490\pm90\pm1820$ & $12\,970\pm90\pm1310$  \\ [\cmsTabSkip]
Data  &  6787  &  8633  &  9931  &  13\,085   \\
\hline
\end{tabular}
}

\end{table}

Table~\ref{tab:systsall} lists the systematic uncertainties in the signal strength after the fit.
The effect of the uncertainties on the signal strength is estimated by a likelihood scan where only one nuisance parameter (or a group of them) is varied at once
while the others are fixed to their nominal postfit values.
The largest experimental uncertainties are from $\tauh$ identification and misidentification, and pileup estimation.
The largest theoretical uncertainties are due to the modelling of top quark \pt in \ttbar processes, \PQb quark fragmentation, and PS modelling (ISR and FSR).

\begin{table}[htp]
\centering
\topcaption{Systematic and statistical uncertainties
determined from the fit to the data in the $\Pe\tauh$ and $\PGm\tauh$ final states, and their combination.
Uncertainties are grouped by their origin: experimental, theoretical, normalization, and extrapolation.
The uncertainties in the measurement in the dilepton final state~\cite{Sirunyan:2018goh} used in the partial width ratio estimate
are also
quoted (column "dileptons"), where the asymmetric extrapolation uncertainties are symmetrized by adding them in quadrature.
As both measurements use the same data, some uncertainties in the $\ell\tauh$ and light dilepton final states are correlated, as shown in the last column.
}
\label{tab:systsall}

\cmsTable{
\begin{tabular}{lrrrrc}
\hline
\multicolumn{1}{c}{Source} & \multicolumn{5}{c}{Uncertainty [\%]} \\
                           &  $\Pe\tauh$        & $\PGm\tauh$    &     Combined  & Dileptons & Correlation \\

\multicolumn{4}{l}{Experimental uncertainties} & & \\
  \hspace*{4mm} $\tauh$ jet identification          &  4.7 &  4.5 &  4.5 & \NA & 0 \\
  \hspace*{4mm} $\tauh$ jet misidentification       &  2.2 &  2.3 &  2.3 & \NA & 0 \\
  \hspace*{4mm}   Pileup			                &  2.5 &  2.2 &  2.3 & 0.1 & 1 \\
  \hspace*{4mm} Lepton identification and isolation &  1.8 &  1.1 &  1.2 & 2.0 & 1 \\

  \hspace*{4mm} \PQb tagging efficiency   &  1.1 &  1.2 &  0.9 & 0.4 & 1 \\
  \hspace*{4mm}     $\tauh$ energy scale  &  0.7 &  0.8 &  0.8 & \NA & 0 \\

  \hspace*{4mm} Trigger efficiency        & 2.3 & 0.6 & 0.7 & 0.3 & 0 \\

  \hspace*{4mm}    Drell--Yan background  &  0.4 &  0.4 &  0.6 & 0.9 & 1 \\
  \hspace*{4mm}        \ttbar background  &  1.0 &  0.8 &  0.6 & 0.2 & 0 \\
  \hspace*{4mm}            tW background  &  0.6 &  0.5 &  0.5 & 1.1 & 1 \\
  \hspace*{4mm}    {\PW}+jets background  &  0.1 &  0.4 &  0.5 & 0.2 & 0 \\
  \hspace*{4mm}      Multijet background  &  0.1 &  0.5 &  0.4 & $<$0.1 & 0 \\
  \hspace*{4mm}         Jet energy scale  &  0.1 &  0.2 &  0.4 & 0.4 & 1 \\

  \hspace*{4mm}    Jet energy resolution  &  0.6    &  0.3    & 0.1    & 0.4 & 1 \\
  \hspace*{4mm} Electron momentum scale   &  0.1    &  0.1    & 0.1    & 0.1 & 1 \\
  \hspace*{4mm}    Muon momentum scale    &  0.1    &  0.1    & 0.1    & 0.1 & 1 \\
  \hspace*{4mm}   Diboson  background     &  $<$0.1 &  $<$0.1 & $<$0.1 & 0.2 & 1 \\

\multicolumn{4}{l}{Theoretical uncertainties} & & \\
  \hspace*{4mm}           \PQb fragmentation  &  2.3 &  2.0 &  2.4 & 0.7 & 1 \\
  \hspace*{4mm}  Top quark $\pt$ modelling  &  2.7 &  2.3 &  2.2 & 0.5 & 1 \\
  \hspace*{4mm}          \ttbar FSR scale  &  1.7 &  1.9 &  1.7 & 0.8 & 1 \\
  \hspace*{4mm}              tW FSR scale  & $<$0.1 & $<$0.1 & $<$0.1 & 0.1 & 1 \\
  \hspace*{4mm}          \ttbar ISR scale  &  1.7 &  1.6 &  1.5 & 0.4 & 1 \\
  \hspace*{4mm}              tW ISR scale  & $<$0.1 & $<$0.1 & $<$0.1 & 0.1 & 1 \\
  \hspace*{4mm}          \ttbar ME scale   &  1.1 &  1.2 &  1.1 & 0.2 & 1 \\
  \hspace*{4mm}              tW ME scale   & $<$0.1 & $<$0.1 & $<$0.1 & 0.2 & 1 \\
  \hspace*{4mm}      Drell--Yan ME scale   & $<$0.1 & $<$0.1 & $<$0.1 & 0.1 & 1 \\
  \hspace*{4mm} Semileptonic \PQb hadron branching fraction  &  0.8 &  0.6 &  0.7 & 0.1 & 1 \\
  \hspace*{4mm}          Underlying event  &  0.5 &  0.5 &  0.6 & 0.3 & 1 \\
  \hspace*{4mm}            ME-PS matching  &  0.4 &  0.4 &  0.5 & 0.2 & 1 \\

  \hspace*{4mm} Colour reconnection & $<$0.1 & $<$0.1 & $<$0.1 & 0.3 & 1 \\
  \hspace*{4mm} PDFs                &   1.5  &   1.5  &   1.6  & 1.1 & 1 \\

\multicolumn{4}{l}{Normalization uncertainties} & & \\
  \hspace*{4mm}       Statistical      & 1.4 & 1.1 & 0.9 & 0.2 & 0 \\
  \hspace*{4mm}    MC statistical      & 2.0 & 1.6 & 1.6 & 1.1 & 0 \\
  \hspace*{4mm}  Integrated luminosity & 2.5 & 2.5 & 2.5 & 2.5 & 1 \\

\multicolumn{4}{l}{Extrapolation uncertainties} & & \\
  \hspace*{4mm} \ttbar ME scale               & 0.3 & 0.4 & 0.3 & 0.3    & 0 \\
  \hspace*{4mm} PDFs                          & 1.2 & 1.4 & 1.3 & 1.0    & 0 \\
  \hspace*{4mm} Top quark $\pt$ modelling     & 1.0 & 1.1 & 1.1 & 0.5    & 0 \\
  \hspace*{4mm} \ttbar ISR scale              & 0.5 & 0.3 & 0.3 & 0.1    & 0 \\
  \hspace*{4mm} \ttbar FSR scale              & 1.9 & 2.0 & 1.9 & 0.1    & 0 \\
  \hspace*{4mm} Underlying event              & 0.3 & 0.2 & 0.2 & $<$0.1 & 0 \\
\hline
\end{tabular}
}
\end{table}

The fiducial cross section for the production of \ttbar events is extracted from the acceptance region of
kinematic phase space defined by the selection criteria described earlier.
The estimate of the fiducial cross section includes the branching fractions of the final states, trigger, lepton identification and isolation, and the overall reconstruction efficiency.
The cross sections in the fiducial phase space for the individual $\Pe\tauh$ and $\PGm\tauh$ final states, as well as the $\ell\tauh$ combined final state,
are measured from the PLR fit to be:
\begin{align}
	\sigma^{\mathrm{fid}}_{\ttbar}(\Pe\tauh)	&= 133.2 \pm 1.9\stat \pm  10.9\syst \pm  3.3\lum\unit{pb}, \\
	\sigma^{\mathrm{fid}}_{\ttbar}(\PGm\tauh)	&= 135.2 \pm 1.5\stat \pm   9.9\syst \pm  3.4\lum\unit{pb}, \\
	\sigma^{\mathrm{fid}}_{\ttbar}(\ell\tauh)	&= 134.5 \pm 1.2\stat \pm   9.5\syst \pm  3.4\lum\unit{pb}.
\end{align}
The acceptance $A_{\ttbar}$ is the fraction of signal events in the fiducial phase space, and it is determined with respect to all signal events in the nominal \ttbar simulation.
It includes kinematic selection cuts and is evaluated for the different signal final states as:
\begin{align}
	A_{\ttbar}(\Pe\tauh)  &= 0.1687 \pm 0.0004\stat \pm 0.0060\syst, \\
	A_{\ttbar}(\PGm\tauh) &= 0.1756 \pm 0.0004\stat \pm 0.0065\syst, \\
	A_{\ttbar}(\ell\tauh) &= 0.1722 \pm 0.0003\stat \pm 0.0062\syst,
\end{align}
where the systematic uncertainties include the uncertainties of the modelling as described in Section~\ref{sec:systematics} and listed as ``Extrapolation uncertainties'' in Table~\ref{tab:systsall}.

\begin{figure}[htp]
\centering
\includegraphics[width=0.6\textwidth]{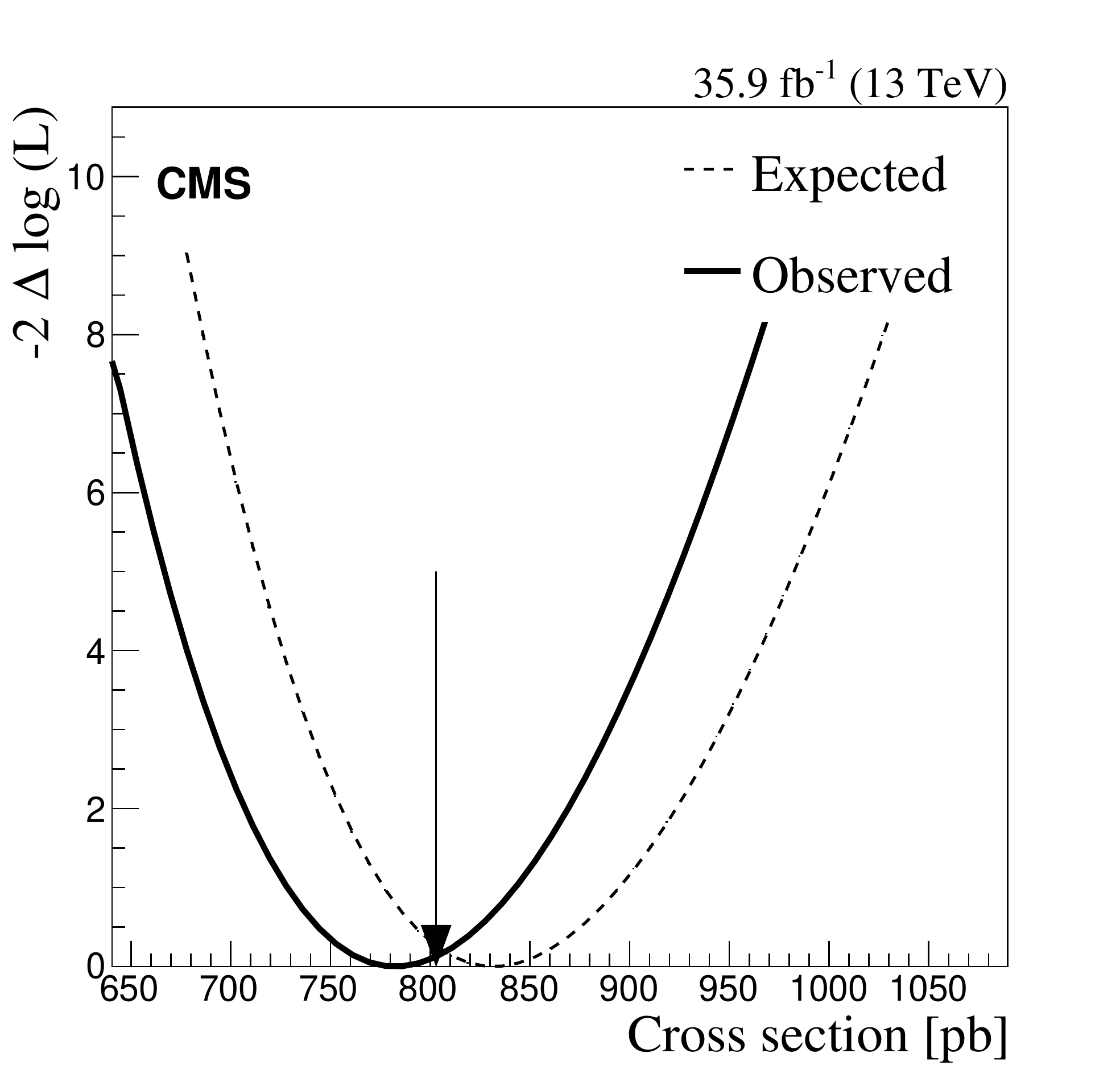}
\caption{
The expected and observed dependence of the likelihood on the total \ttbar cross section $\sigma_{\ttbar}$.
It is derived from the fiducial phase space by a simple extrapolation.
The arrow points at the cross section measured in the light dilepton final state.
The goodness of the fit determined with a Kolmogorov--Smirnov method yields a $p$ value of 0.24.
}
\label{fig:scans}
\end{figure}

The cross section values in the full phase space are obtained from the extrapolation of the fiducial cross sections
using the acceptances $A_{\ttbar}$ estimated from the simulation:
\begin{align}
\sigma_{\ttbar}(\Pe\tauh)	&= 789 \pm 11\stat \pm 71\syst \pm 20\lum\unit{pb}, \\
\sigma_{\ttbar}(\PGm\tauh)	&= 770 \pm  8\stat \pm 63\syst \pm 20\lum\unit{pb}, \\
\sigma_{\ttbar}(\ell\tauh)	&= 781 \pm  7\stat \pm 62\syst \pm 20\lum\unit{pb}.
\end{align}
The expected and observed dependence of the likelihood on the cross section in the full phase space
in the $\ell\tauh$ combined final state
are shown in Fig.~\ref{fig:scans}.
The result of the fit is
consistent with the predicted SM \ttbar production cross section of
$832^{+20}_{-29}\,\mathrm{(scale)}\pm35\,\mathrm{(PDF+}\alpS)$\unit{pb}~\cite{Czakon:2011xx}.
Using simulated \ttbar samples with different $m_{\PQt}$ values, we find that the cross section changes by 1.5\% per $\Delta m_{\PQt} = 1$\GeV.

The ratio of the cross section in the $\ell\tauh$ final state divided by the cross section measured in the dilepton final state in the same data-taking period~\cite{Sirunyan:2018goh} yields a value of
$R_{\ell\tauh/\ell\ell}=0.973 \pm 0.009\stat \pm 0.066\syst$,
consistent with unity as expected from lepton flavour universality.
The relative systematic uncertainty in the ratio is 6.8\%.
About 5\% comes from the uncertainties in the \tauh identification (4.5\%) and misidentification probability in \ttbar events (2.3\%).
The rest comes from the other uncorrelated uncertainties in the ratio and the treatment of the correlated uncertainties in the calculation of the ratio.
In particular, a small contribution comes from the uncertainties in the extrapolation to the full phase space that are considered uncorrelated because the two measurements
extrapolate from different fiducial phase spaces.
Also, the triggers are not the same.

The measurement also provides an estimate of the ratio of the partial to the total width of the top quark decay,
$R_{\Gamma}=\Gamma(\PQt\to\PGt\nu_{\PGt}\PQb)/\Gamma_{\text{total}}$.
The ratio is calculated as
$R_{\Gamma}=\sigma_{\ttbar}(\ell\tauh) \mathcal{B}(\PW\to\PGt\nu_\PGt)/\sigma_{\ttbar}(\ell\ell)$,
where
the cross section measured in the $\ell\tauh$ final state is multiplied by the branching fraction
$\mathcal{B}(\PW\to\PGt\nu_\PGt)$
and divided by the inclusive \ttbar cross section measured in the dilepton final state~\cite{Sirunyan:2018goh}.
The \PW boson branching fraction $\mathcal{B}(\PW\to\PGt\nu_\PGt)$ that is included in the signal acceptance is cancelled out in the multiplication.
Since both measurements are performed in the same data-taking period with the same reconstruction algorithms,
the uncertainty in the ratio includes the correlations between common sources of uncertainties as indicated in Table~\ref{tab:systsall}.
The estimate yields the value
$R_{\Gamma}=0.1050 \pm 0.0009\stat \pm 0.0071\syst$,
improving over
the previous measurements~\cite{Aad:2015dya,Aaltonen:2014hua,PDG2018}.
The result is dominated by the systematic uncertainty and it is consistent with the SM value of $0.1083\pm0.0002$~\cite{PDG2018}.
While in Ref.~\cite{Aad:2015dya} the partial width is evaluated for hadronic decays of $\PGt$ leptons,
here $R_\Gamma$ is measured for all $\PGt$ decays by using the $\mathcal{B}(\PGt\to\tauh\PGn_\PGt) = 64.8 \pm 0.1 \%$ branching fraction~\cite{PDG2018}.

\section{Summary}
\label{sec:summary}

A measurement of the top quark pair production cross section
in the $\ttbar\to (\ell \PGn_{\ell}) (\tauh  \PGn_{\PGt}) \bbbar$ channel,
where $\ell$ is either an electron or a muon,
is performed by CMS in proton-proton collisions at LHC,
using a data sample corresponding to an integrated luminosity of 35.9\fbinv obtained at $\sqrt{s} = 13\TeV$.
Events are selected by requiring the presence of an electron or a muon,
and at least three jets, of which at least one is \PQb tagged and one is identified as a
\PGt lepton decaying to hadrons ($\tauh$).
The largest background contribution arises from \ttbar lepton+jets events, $\ttbar\to (\ell\PGn_{\ell})(\PQq\PAQq')\bbbar$, where one jet is misidentified as the $\tauh$.
The background contribution is constrained in a fit to the distribution of the transverse mass of the light lepton and missing transverse momentum system
in two event categories,
constructed according to the kinematic properties of the jets in the \ttbar lepton+jets final state.
The signal enters as a free parameter without constraining the kinematic properties of the $\PGt$ lepton.
Assuming a top quark mass of 172.5\GeV, the measured total \ttbar cross section
$\sigma_{\ttbar}(\ell\tauh)	= 781 \pm  7\stat \pm 62\syst \pm 20\lum\unit{pb}$ is
in agreement with the standard model expectation.
This is the first measurement of the \ttbar production cross section in proton-proton collisions at $\sqrt{s}=13\TeV$ that explicitly includes hadronically decaying $\PGt$ leptons,
and it improves the relative precision with respect to the 7 and 8\TeV results~\cite{Chatrchyan:2012vs,Khachatryan:2014loa}.
The higher precision is achieved through a shape fit to the kinematic distributions of the events, thus better constraining the backgrounds.
The measurement of the ratio of the cross section in the $\ell\tauh$ final state to the light dilepton cross section~\cite{Sirunyan:2018goh} yields a value of
$R_{\ell\tauh/\ell\ell}=0.973 \pm 0.009\stat \pm 0.066\syst$, consistent with lepton universality.
The ratio of the partial to the total width of the top quark
$\Gamma(\PQt\to\PGt\PGn_{\PGt}\PQb)/\Gamma_{\text{total}}=0.1050 \pm 0.0009\stat \pm 0.0071\syst$
is measured with respect to the \ttbar inclusive cross section extrapolated from the light dilepton final state,
improving the precision over the previous measurements~\cite{Aaltonen:2014hua,Aad:2015dya}.

\begin{acknowledgments}
We congratulate our colleagues in the CERN accelerator departments for the excellent performance of the LHC and thank the technical and administrative staffs at CERN and at other CMS institutes for their contributions to the success of the CMS effort. In addition, we gratefully acknowledge the computing centres and personnel of the Worldwide LHC Computing Grid for delivering so effectively the computing infrastructure essential to our analyses. Finally, we acknowledge the enduring support for the construction and operation of the LHC and the CMS detector provided by the following funding agencies: BMBWF and FWF (Austria); FNRS and FWO (Belgium); CNPq, CAPES, FAPERJ, FAPERGS, and FAPESP (Brazil); MES (Bulgaria); CERN; CAS, MoST, and NSFC (China); COLCIENCIAS (Colombia); MSES and CSF (Croatia); RPF (Cyprus); SENESCYT (Ecuador); MoER, ERC IUT, PUT and ERDF (Estonia); Academy of Finland, MEC, and HIP (Finland); CEA and CNRS/IN2P3 (France); BMBF, DFG, and HGF (Germany); GSRT (Greece); NKFIA (Hungary); DAE and DST (India); IPM (Iran); SFI (Ireland); INFN (Italy); MSIP and NRF (Republic of Korea); MES (Latvia); LAS (Lithuania); MOE and UM (Malaysia); BUAP, CINVESTAV, CONACYT, LNS, SEP, and UASLP-FAI (Mexico); MOS (Montenegro); MBIE (New Zealand); PAEC (Pakistan); MSHE and NSC (Poland); FCT (Portugal); JINR (Dubna); MON, RosAtom, RAS, RFBR, and NRC KI (Russia); MESTD (Serbia); SEIDI, CPAN, PCTI, and FEDER (Spain); MOSTR (Sri Lanka); Swiss Funding Agencies (Switzerland); MST (Taipei); ThEPCenter, IPST, STAR, and NSTDA (Thailand); TUBITAK and TAEK (Turkey); NASU (Ukraine); STFC (United Kingdom); DOE and NSF (USA).

\hyphenation{Rachada-pisek} Individuals have received support from the Marie-Curie programme and the European Research Council and Horizon 2020 Grant, contract Nos.\ 675440, 752730, and 765710 (European Union); the Leventis Foundation; the A.P.\ Sloan Foundation; the Alexander von Humboldt Foundation; the Belgian Federal Science Policy Office; the Fonds pour la Formation \`a la Recherche dans l'Industrie et dans l'Agriculture (FRIA-Belgium); the Agentschap voor Innovatie door Wetenschap en Technologie (IWT-Belgium); the F.R.S.-FNRS and FWO (Belgium) under the ``Excellence of Science -- EOS" -- be.h project n.\ 30820817; the Beijing Municipal Science \& Technology Commission, No. Z181100004218003; the Ministry of Education, Youth and Sports (MEYS) of the Czech Republic; the Deutsche Forschungsgemeinschaft (DFG) under Germany’s Excellence Strategy -- EXC 2121 ``Quantum Universe" -- 390833306; the Lend\"ulet (``Momentum") Programme and the J\'anos Bolyai Research Scholarship of the Hungarian Academy of Sciences, the New National Excellence Program \'UNKP, the NKFIA research grants 123842, 123959, 124845, 124850, 125105, 128713, 128786, and 129058 (Hungary); the Council of Science and Industrial Research, India; the HOMING PLUS programme of the Foundation for Polish Science, cofinanced from European Union, Regional Development Fund, the Mobility Plus programme of the Ministry of Science and Higher Education, the National Science Center (Poland), contracts Harmonia 2014/14/M/ST2/00428, Opus 2014/13/B/ST2/02543, 2014/15/B/ST2/03998, and 2015/19/B/ST2/02861, Sonata-bis 2012/07/E/ST2/01406; the National Priorities Research Program by Qatar National Research Fund; the Ministry of Science and Education, grant no. 3.2989.2017 (Russia); the Programa Estatal de Fomento de la Investigaci{\'o}n Cient{\'i}fica y T{\'e}cnica de Excelencia Mar\'{\i}a de Maeztu, grant MDM-2015-0509 and the Programa Severo Ochoa del Principado de Asturias; the Thalis and Aristeia programmes cofinanced by EU-ESF and the Greek NSRF; the Rachadapisek Sompot Fund for Postdoctoral Fellowship, Chulalongkorn University and the Chulalongkorn Academic into Its 2nd Century Project Advancement Project (Thailand); the Nvidia Corporation; the Welch Foundation, contract C-1845; and the Weston Havens Foundation (USA).
\end{acknowledgments}

\bibliography{auto_generated}

\cleardoublepage \appendix\section{The CMS Collaboration \label{app:collab}}\begin{sloppypar}\hyphenpenalty=5000\widowpenalty=500\clubpenalty=5000\vskip\cmsinstskip
\textbf{Yerevan Physics Institute, Yerevan, Armenia}\\*[0pt]
A.M.~Sirunyan$^{\textrm{\dag}}$, A.~Tumasyan
\vskip\cmsinstskip
\textbf{Institut f\"{u}r Hochenergiephysik, Wien, Austria}\\*[0pt]
W.~Adam, F.~Ambrogi, T.~Bergauer, J.~Brandstetter, M.~Dragicevic, J.~Er\"{o}, A.~Escalante~Del~Valle, M.~Flechl, R.~Fr\"{u}hwirth\cmsAuthorMark{1}, M.~Jeitler\cmsAuthorMark{1}, N.~Krammer, I.~Kr\"{a}tschmer, D.~Liko, T.~Madlener, I.~Mikulec, N.~Rad, J.~Schieck\cmsAuthorMark{1}, R.~Sch\"{o}fbeck, M.~Spanring, D.~Spitzbart, W.~Waltenberger, C.-E.~Wulz\cmsAuthorMark{1}, M.~Zarucki
% ...
\vskip\cmsinstskip
\dag: Deceased\\
1:  Also at Vienna University of Technology, Vienna, Austria\\
% ...
\end{sloppypar}
\end{document}